\documentclass[10pt,aps,prl,twocolumn,notitlepage]{revtex4-1}
\usepackage{amsmath}
\usepackage{amssymb}
\usepackage{graphicx}
\usepackage{float}
\usepackage{xcolor}
\colorlet{RED}{red}
\usepackage{tabularray}
\usepackage{eso-pic}
\usepackage{bibunits}
\usepackage[hidelinks,hypertexnames=false]{hyperref}
\renewcommand{\bibsection}{\section*{References}}

\begin{document}

\title{Thermally Hysteretic Room-Temperature Nonlinear Hall Effect in Bulk EuTe$_4$}
\author{Yang Fu,$^{1,2}$ Xingyu Lv,$^{2}$ Shangjie Tian,$^{2}$ Ying Ma,$^{3}$ Xiao Zhang,$^{3,*}$ Cedomir Petrovic,$^{4,1,6}$ and Hechang Lei$^{2,5,*}$}
\affiliation{
$^{1}$Center for High Pressure Science and Technology Advanced Research (HPSTAR), Beijing 100193, China\\
$^{2}$School of Physics and Key Laboratory of Quantum State Construction and Manipulation (Ministry of Education), Renmin University of China, Beijing 100872, China\\
$^{3}$State Key Laboratory of Information Photonics and Optical Communications $\&$ School of Physical Science and Technology, Beijing University of Posts and Telecommunications, Beijing 100876, China\\
$^{4}$Shanghai Key Laboratory of Material Frontiers Research in Extreme Environments (MFree), Shanghai Advanced Research in Physical Sciences (SHARPS), Shanghai 201203, China\\
$^{5}$Xiongan Future Industrial Institute, Renmin University of China, Xiongan 071700, China\\
$^{6}$Tsung-Dao Lee Institute, Shanghai Jiao Tong University, Shanghai 200240, China
}

\date{\today}
\begin{abstract}
The nonlinear Hall effect (NLHE) is a novel transport phenomena and shows great potential applications of future electronics.
However, the realization of a giant and tunable NLHE in bulk materials at room temperature remains elusive.
Here, we report a NLHE in bulk EuTe$_4$ single crystals, which persists up to 400 K and exhibits pronounced thermal hysteresis behaviors.
The ratio of nonlinear Hall voltages during warming and cooling processes is about 4.7 at 340 K.
Such phenomena could be related to the inversion-symmetry-broken charge density wave state in EuTe$_4$.
These findings demonstrate bulk EuTe$_4$ as a promising platform for tunable nonlinear electronic devices at room temperature.
\end{abstract}
\maketitle

The discovery of nonlinear Hall effect (NLHE), characterized by a transverse voltage quadratic in the applied current, has opened new avenues for probing the symmetry and topological properties of quantum materials with time-reversal symmetry, and developing novel electronic devices \cite{Du2021NLHReview,Ortix2021NLHReview,Bandyopadhyay2024NLHReview,SuarezRodriguez2025NonlinearReview,Li2021HallReview,Wang2025NLHReview,Wu2023MoireReview,Wu2026QGTHzReview}.
Guided by the theoretical predictions \cite{Sodemann}, a Berry curvature dipole (BCD)-induced NLHE was first observed in semimetal WTe$_2$ \cite{MaQ,KangK}.
However, the observed relaxation-time dependence can not be fully explained by the BCD mechanism alone, pointing to additional contributions from disorder-related processes such as skew scattering.
Scaling analysis is developed to separate intrinsic and extrinsic contributions \cite{Du2019}.
Later on, further experiments identify that quantum metric can also lead to the NLHE \cite{Gao2023,Wang2023}.

The second-order NLHE have been observed in various systems, including topological semimetals and insulators Ta(Nb)IrTe$_4$ and Bi$_2$Se$_3$ \cite{MaQ,KangK,Tiwari2021MoTe2,KumarD,He2021Bi2Se3},
interfacial electron gases systems like LaAlO$_3$/SrTiO$_3$ and CaZrO$_3$/KTaO$_3$ \cite{Lesne2023,Zhang2026KTaO3,Dash2025ZnO},
semiconductors without inversion symmetry such as Te and CdGeAs$_2$ \cite{Cheng2024Te,Liu2025TeRectifier,Lee2026CdGeAs2,Kim2024TeGate},
and charge density wave (CDW) material 1$T$-TaS$_2$ \cite{Alam2025TaS2,Wu2025TaS2,Zhao2026TaS2THz}.
Moreover, the NLHE is of high interest for potential applications including radiofrequency/microwave rectification and terahertz detection \cite{KumarD,Cheng2024Te,Hu2023CoTe2,Xi2025TaIrTe4,Min2023BaMnSb2}%
, frequency multiplication and mixing \cite{He2021Bi2Se3,Min2023BaMnSb2,Lee2026CdGeAs2,Lee2025EuSc2Te4,Makushko2024Bi}%
, and Hall-based sensing \cite{Liu2025Strain,Chen2025Thermometer}.
However, many of reported large NLHEs are observed mainly at low temperatures or primarily in micro- and nanoscale devices.
The realization of a sizable and tunable NLHE in bulk materials at room temperature remains elusive.

The quasi-two-dimensional rare-earth telluride EuTe$_4$ hosts a CDW state above the room temperature \cite{WuD,LiuQM2024,LvBQ2022}.
This CDW state breaks inversion symmetry and forms a quasi-two-dimensional (quasi-2D) polarized structure, as evidenced by room-temperature second-harmonic generation (SHG) measurements \cite{LiuQM2024}.
The broken inversion symmetry implies that a CDW-driven NLHE may exist in EuTe$_4$.
Moreover, the CDW state exhibits thermal hysteresis behaviors and can be modulated by optical and electrical stimuli \cite{WuD,Venturini2026,Lv2026,LvBQ}.
This hysteresis suggests that the NLHE might be tuned and applied for thermal memory devices \cite{Venturini2026,Lv2026}.
Here, we report a large room-temperature NLHE in bulk EuTe$_4$ single crystals that persists up to 400 K.
Systematic transport measurements show that the nonlinear Hall signal closely tracks the thermal hysteresis of the CDW state.
Scaling analysis reveals two distinct regimes: at high temperatures, the NLHE follows the conventional scaling relation with longitudinal conductivity, whereas at lower temperatures it progressively deviates from this behavior. %

EuTe$_4$ possesses a quasi-2D crystal structure. It is composed of a Te-[EuTe]-Te-Te-[EuTe] stacking sequence along the $c$ axis [Fig. 1(a)]  \cite{WuD}, featuring both single- and double-layer Te square nets at high temperature.
When entering the CDW state at lower temperature, the Te square lattices undergo structural distortion, forming Te atom trimers [Fig. 1(b)].
Crucially, the distortion and polarization of these Te trimers intrinsically break the inversion symmetry, providing the fundamental prerequisite for generating the NLHE.
The intralayer polarization direction determined by the direction of structural distortion of Te trimers may be different in the single and double Te layers, which are schematically represented by the arrows in Fig. 1(a).
It is noted that the polarization sequence between different layers is not rigidly fixed, potentially giving rise to up to 64 distinct polarization configurations \cite{LiuQM2024}.
In a macroscopic sample, the stacking of different inter-layer polarization configuration lead to the formation of out-of-plane CDW domains, while different orientations of the intra-layer trimer distortions give rise to in-plane CDW domains. Despite this complexity, the macroscopic nonlinear Hall responses can be considered as a simple superposition of the contributions from multiple domains.
For a domain-free Te layer, due to the constrains of crystallographic symmetry, the nonlinear transport behaviors are governed by the following formula (\hyperref[sec:note1]{Supplemental Note~1}),

\[\begin{bmatrix} J_a^{2\omega}\\J_b^{2\omega}\end{bmatrix}=\begin{bmatrix}  \chi_{aaa}&  \chi_{abb}&  0 \\  0&  0&  \chi_{bba} \end{bmatrix}\left[\begin{array}{c}(E_a^{\omega})^2 \\(E_b^{\omega})^2 \\2E_a^{\omega}E_b^{\omega}\end{array}\right]\]

where $J_a^{2\omega}$ and $J_b^{2\omega}$ are the nonlinear current densities along the $a$- and $b$-axes, $E_a^{\omega}$ and $E_b^{\omega}$ represent the applied alternating electric field along the $a$- and $b$-axes, and $\chi_{aaa}$, $\chi_{abb}$ and $\chi_{bba}$ are the independent non-zero elements of the nonlinear conductivity tensor.
According to these symmetry constraints, an $E_b^{\omega}$ should produce a purely transverse nonlinear current $J_a^{2\omega} = \chi_{abb} (E_b^{\omega})^2$. Conversely, an $E_a^{\omega}$ should generate a purely longitudinal nonlinear current $J_a^{2\omega} = \chi_{aaa} (E_a^{\omega})^2$.
When electric field deviates from the principal crystallographic directions, there will be a finite nonlinear signal along the $b$-axis.
To experimentally probe these nonlinear signals, we applied an $I^{\omega}$ along specific crystallographic axes and simultaneously measured the second-harmonic transverse and longitudinal voltages ($V^{2\omega}$)  using lock-in amplifiers.
Based on the Ohm's law $\mathbf{E}^{2\omega} = \boldsymbol{\rho} \cdot \mathbf{J}^{2\omega}$, where $\boldsymbol{\rho}$ is the linear resistivity tensor, the nonlinear conductivities ($\chi_{abb}$ and $\chi_{aaa}$) can be rigorously extracted from the measured  $I^{\omega}$ and $V^{2\omega}$ (\hyperref[sec:note2]{Supplemental Note~2}).

EuTe$_4$ single crystals were synthesized using the self-flux method (\hyperref[sec:note3]{Supplemental Note~3}). The X-ray diffraction (XRD) pattern and optical image [Fig. 1(c)] confirm that the crystal surface is along the $ab$-plane with high quality. Fig. 1(d) shows the temperature-dependent zero-field longitudinal resistivity $\rho_{bb}(T)$ along the $b$-axis of EuTe$_4$ single crystal. The $\rho_{bb}(T)$ exhibits a thermal hysteresis between 200 K and 400 K, consistent with previously reported results \cite{LvBQ2022, ZhangQQ, LiuQM2024}. It indicates the presence of a CDW transition.
Between 74 K -- 200 K, the $\rho_{bb}(T)$ curve exhibits a metallic-like behavior without obvious hysteresis. With lowering temperature further, EuTe$_4$ shows a semiconducting behavior again.
This complex temperature dependence could be understood by assuming the existence of metallic CDW domain boundaries. However, these metallic pathways occupy a very small volume fraction, and thus do not pull down the bulk resistivity by orders of magnitude. Below 80~K, the metallic domain boundaries gradually become connected upon warming, and the high-resistance CDW phase dominates, leading to semiconducting behavior.
Between 80 K and 360 K, the connected domain boundaries shunt the resistivity of CDW domains; as temperature rises, their metallic resistance increases, giving rise to metallic behavior.
Above 360 K, the domain boundaries network expands quickly on warming process, leading to the drop near 380 K. On cooling process, the domain boundaries contract sluggishly, causing the steep rise of resistivity near 320 K and forming the thermal hysteresis of resistivity.
When cooled to 100 K --200 K, the residual CDW regions surviving at 400 K \cite{LvBQ2022,ZhangQQ} act as nucleation centers to form a similar domain configuration of the warming process, restoring nearly identical longitudinal resistivity.

\begin{figure}[t]
\centering\includegraphics[width=\columnwidth]{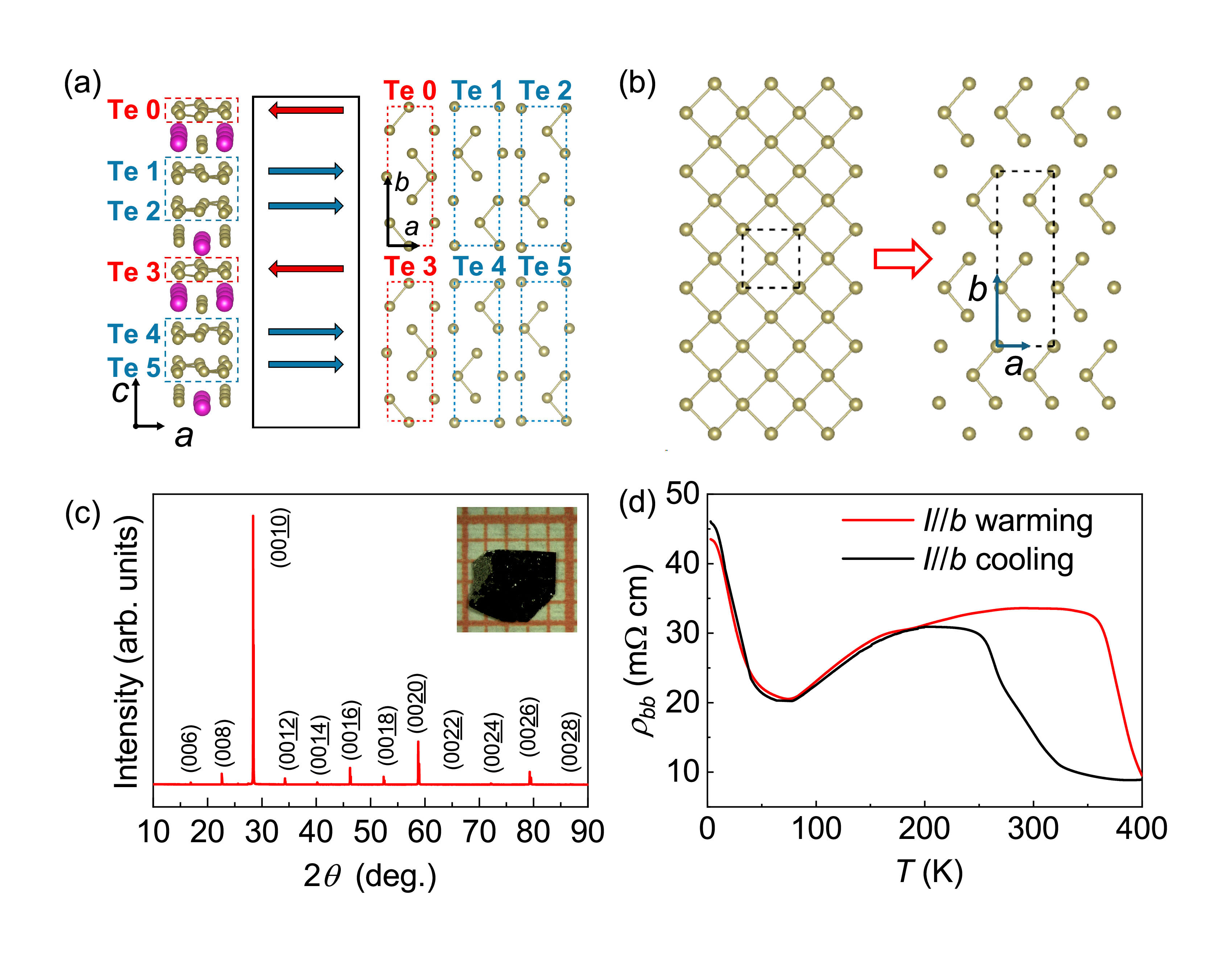}
\caption{
	(a) Stacking configuration of the CDW states along the $c$-axis (left) and atomic positions in each Te layers along the $ab$-plane (right). The arrows schematically indicate the intralayer polarization directions within the single (Te 0 and Te 3) and double (Te 1/Te 2 and Te 4/Te 5) Te layers.
	(b) Schematic of the structural distortion in one Te layer. The left panel shows the pristine Te square lattice, and the right panel illustrates the distorted lattice in the CDW state, characterized by the formation of Te trimers.
	(c) XRD pattern of a EuTe$_4$ single crystal. Inset: an optical image of typical EuTe$_4$ single crystal.
	(d) Temperature dependence of zero-field $\rho_{bb}(T)$ with current applied along the $b$-axis.}
\end{figure}

To probe the nonlinear transport behaviors, we measured the second-harmonic voltage $V^{2\omega}$ as a function of the applied AC current $I^{\omega}$ during the warming process.
As depicted in Figs. 2(a) and 2(b), all $V^{2\omega}$ signals display a quadratic dependence on $I^{\omega}$ at 100 K and 300 K, which confirms the second-order nonlinear nature of the response.
Crucially, the $V^{2\omega}$ components follow the symmetry constraints at 300 K.
For current applied along the $b$-axis $I_b^\omega$ (sample 1) [inset of Fig. 2(a)], the transverse voltage $V_{abb}^{2\omega}$ is much larger than the longitudinal one $V_{bbb}^{2\omega}$ ($|V_{abb}^{2\omega}| \gg |V_{bbb}^{2\omega}|$).
Conversely, for current along the $a$-axis $I_a^\omega$ (sample 2) [inset of Fig. 2(b)], the longitudinal voltage $V_{aaa}^{2\omega}$ is much larger than the transverse one $V_{baa}^{2\omega}$ ($|V_{aaa}^{2\omega}| \gg |V_{baa}^{2\omega}|$).
The residual symmetry-forbidden signals (e.g., $V_{bbb}^{2\omega}$ and $V_{baa}^{2\omega}$) likely originate from slight contact misalignments or CDW domains.
Furthermore, while the nonlinear signals exhibit some variation with frequency,
their temperature-dependent behavior remains consistent across different frequencies [\hyperref[sec:note4]{Supplemental Note~4}].
This rules out artifacts such as thermal effects and parasitic capacitive coupling.

\begin{figure}
\centering\includegraphics[width=\columnwidth]{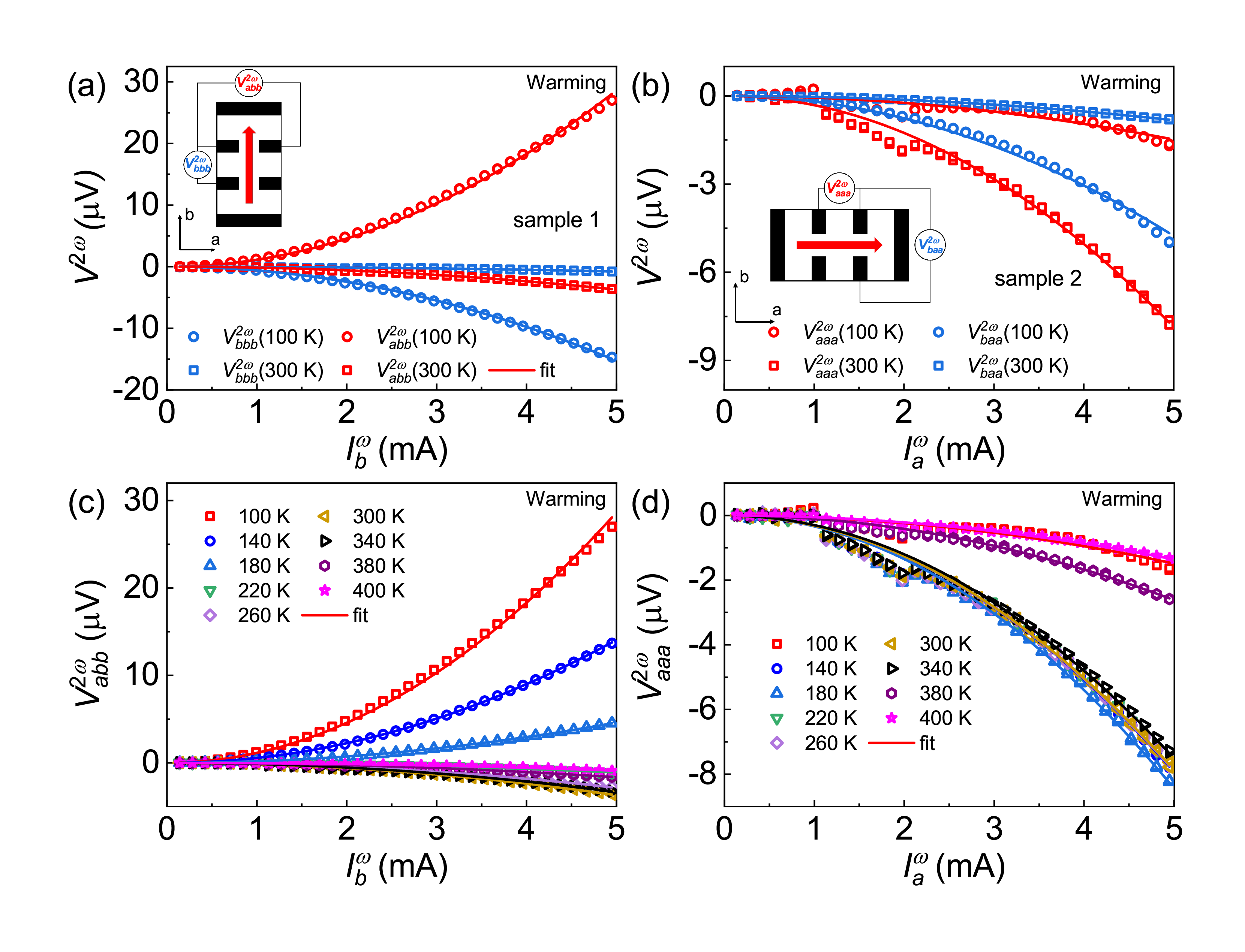}
\caption{
	(a) $V_{abb}^{2\omega}$ and $V_{bbb}^{2\omega}$ versus $I_b^\omega$ at 100 K and 300 K (sample 1).
	(b) $V_{baa}^{2\omega}$ and $V_{aaa}^{2\omega}$ versus $I_a^\omega$ at 100 K and 300 K (sample 2).
	The data for $V_{aaa}^{2\omega}$ at 100 K are multiplied by 0.2 for clarity.
	Insets in (a) and (b) show the schematic figures of measuring configurations, where the red arrows indicate the direction of the applied current.
	(c) $V_{abb}^{2\omega}$ as a function of $I_b^{\omega}$, and (d) $V_{aaa}^{2\omega}$ as a function of $I_a^{\omega}$ at various temperatures. Solid lines in (a)–(d) are quadratic fits.
}
\end{figure}

Figures 2(c) and 2(d) illustrates the temperature dependence of $V_{abb}^{2\omega}$ and $V_{aaa}^{2\omega}$ from 100 K to 400 K during warming process.
The nonlinear voltage signal robustly persists up to 400 K, and the measured $V^{2\omega}$ can be well fitted by a quadratic curve against $I^{\omega}$ across the entire range [\hyperref[sec:note5]{Supplemental Note~5}].
Notably,  the $V_{abb}^{2\omega}$ [Fig. 2(c)] undergoes a distinct sign reversal between 180 K and 220 K while the $V_{aaa}^{2\omega}$ maintains its sign across the measured temperature range [Fig. 2(d)].
This sign change suggests a temperature-driven reorientation of the polarization direction in certain layers [see Fig. 1(a)] or the evolution of intra- and interlayer domains, consistent with synchrotron X-ray diffraction experiments on interlayer CDW stacking \cite{LvBQ,Lv2026} and theoretical predictions of thermally driven redistribution among nearly degenerate polar CDW configurations \cite{Dong2026}.
It is noted that below $T<$ 100 K, the $V^{2\omega}$ clearly deviates from the expected quadratic behavior against $I^{\omega}$.
This deviation is likely attributable to capacitive effects induced by Schottky barriers or a drastic increase in contact resistance ($R_{\rm c}$), which exhibits a pronounced current dependence at low temperatures~[\hyperref[sec:note6]{Supplemental Note~6}].
Furthermore, similar to the warming process, the nonlinear transport effects during the cooling process also exhibits a significant response [\hyperref[sec:note7]{Supplemental Note~7}].

\begin{figure}
	\centering\includegraphics[width=\columnwidth]{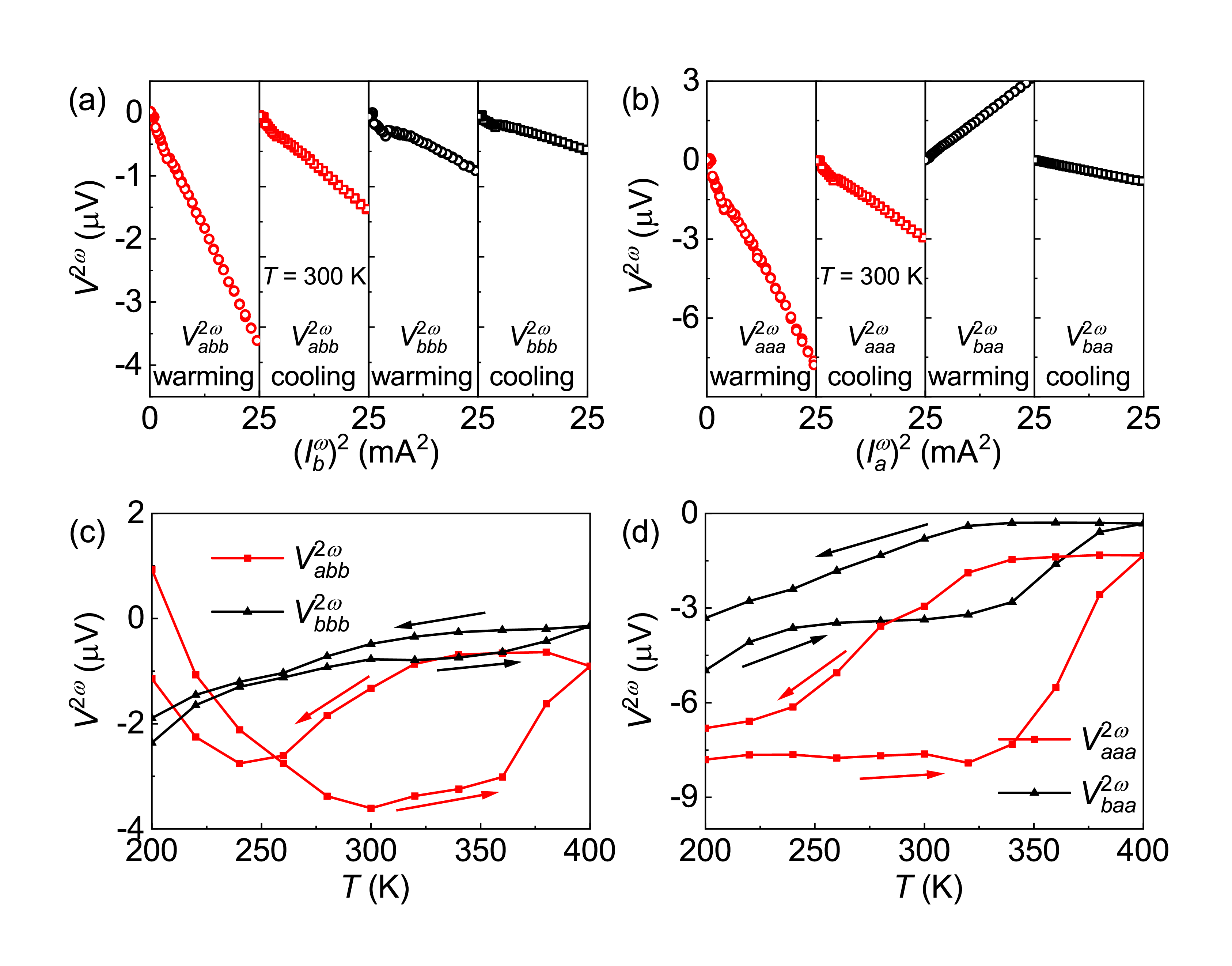}
	\caption{Thermal hysteresis of the nonlinear transport response.
	(a) $V_{abb}^{2\omega}$ and $V_{bbb}^{2\omega}$, and (b) $V_{aaa}^{2\omega}$ and $V_{baa}^{2\omega}$ versus $(I^{\omega})^2$ measured at 300 K during warming and cooling processes.
	In (a) and (b), the $(I^{\omega})^2$ spans from 0 $\mathrm{mA^2}$ to 25 $\mathrm{mA^2}$.
	Temperature dependence of (c) $V_{abb}^{2\omega}$ and $V_{bbb}^{2\omega}$, and (d) $V_{baa}^{2\omega}$ and $V_{aaa}^{2\omega}$ during a thermal cycle.
	Arrows in (c) and (d) indicate the temperature sweep directions.
	}
\end{figure}

To correlate the nonlinear transport effects with the CDW state, we measured the $V^{2\omega}$ as a function of $I^{\omega}$ across a thermal cycle (3 K to 400 K to 3 K).
As shown in Figs. 3(a) and 3(b), the $(I^{\omega})^2$ dependence of $V^{2\omega}$ exhibit distinct slopes between warming and cooling processes at 300 K.
Tracking the temperature dependence [Figs. 3(c) and 3(d)] reveals a clear thermal hysteresis loop in all measured nonlinear voltage components.
Throughout the entire thermal cycle, the symmetry-dictated magnitude relations ($|V_{abb}^{2\omega}| \gg |V_{bbb}^{2\omega}|$ and $|V_{aaa}^{2\omega}| \gg |V_{baa}^{2\omega}|$) are preserved in principle.
Such hysteresis behaviors of $V^{2\omega}(T)$ undoubtedly indicate that the nonlinear transport phenomena are intimately related to the CDW state, which leads to the hysteresis of $\rho_{bb}(T)$ [Fig. 1(d)].
Moreover, the convergence temperature of the hysteresis is close to that of the linear resistivity, further confirming that the observed nonlinear voltage signals are intrinsic.
The ratios $V_{abb}^{2\omega}({\rm warming})/V_{abb}^{2\omega}({\rm cooling})$ and $V_{aaa}^{2\omega}({\rm warming})/V_{aaa}^{2\omega}({\rm cooling})$ reach their maximum values of approximately 4.7 and 4.9 at 340 K, respectively [Figs. 3(c) and 3(d)].

\begin{figure}[tb]
	\centerline{\includegraphics[width=0.57\textwidth]{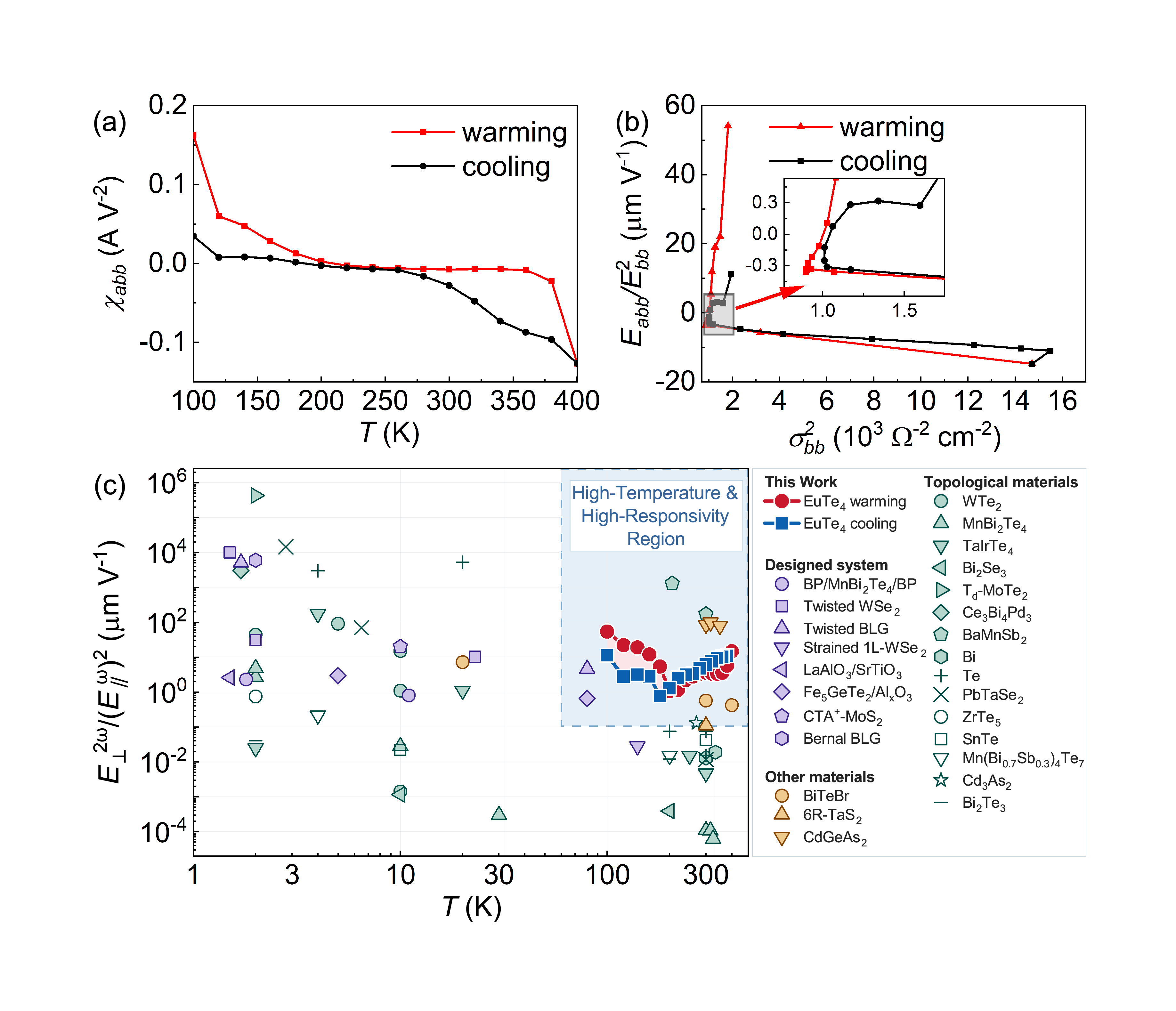}} \vspace*{-0.3cm}
	\caption{
        (a) Temperature dependence of $\chi_{abb}(T)$ measured during the warming and cooling processes between 100 K and 400 K.
        (b) Scaling law analysis of $E_{abb}^{2\omega}/E_{b}^2$ versus $\sigma_{bb}^2$. The inset displays an enlarged view of the data near the origin.
        (c) Comparison of the nonlinear responsivity $|E_{\perp}^{2\omega}|/(E_{\parallel}^{\omega})^2$ as a function of temperature between bulk $\text{EuTe}_4$ and most of reported materials exhibiting the NLHE \cite{MaQ,KangK,Gao2023,Wang2023,KumarD,He2021Bi2Se3,Tiwari2021MoTe2,Dzsaber2021,Huang2023WSe2,Duan2022Graphene,QinMS,Ye2023WTe2BCD,Min2023BaMnSb2,Lesne2023,Makushko2024Bi,Cheng2024Te,Itahashi2022PbTaSe2,Wang2024ZrTe5,Lu2024BiTeBr,Hu2025SnTe,Li2023WTe2Current,Wang2024MnBi4Te7,Liu2024MnBi2Te4,Kim2024TeGate,Zhao2025Cd3As2QMD,Wang2024TaIrTe4Orbital,Cao2025WSe2Scaling,Wu2025TaS2,Zhong2024Fe5GeTe2,Yang2025TaIrTe4Control,Wu2025WTe2Growth,Lee2026CdGeAs2,Shen2025Bi2Te3,Zhou2026MoS2,Chen2026BLG}.
        Red circles and blue squares denote $\text{EuTe}_4$ during warming and cooling, respectively.
        Literature data are classified into topological materials, designed platforms, and other materials (see legend).
        The light-blue shaded region highlights the high-temperature and high-responsivity region.
        Details and conditions are given in \hyperref[sec:note10]{Supplemental Note~10}.
    }
\end{figure}

The temperature-dependent $\chi_{abb}$ and $\chi_{aaa}$ are presented in Figs. 4(a) and \hyperref[sec:note8]{Supplemental Note~8}.
Both $\chi_{abb}$ and $\chi_{aaa}$ exhibit pronounced thermal hysteresis behaviors above about 250 K, similar to those hysteresis of $V^{2\omega}(T)$ and $\rho_{bb}(T)$.
To elucidate the microscopic origin of the NLHE in EuTe$_4$, the scaling law analysis is performed. The nonlinear responsivity is generally expressed as \cite{Du2019, KangK}
$$
\frac{E_{abb}^{2 \omega}}{E_{b}^2}=\xi \sigma_{bb}^2+\eta
$$
The $\xi \sigma_{bb}^2$ term is attributed to skew scattering, while the $\eta$ term originates from the intrinsic contribution (Berry curvature dipole) and side-jump scattering.
As shown in Fig. 4(b), in the high-temperature regime (280 K--400 K for the warming process and 240 K--380 K for the cooling process), the $E_{abb}^{2 \omega}/{E_{b}^2}$ and $\sigma_{bb}^2$ exhibits a clear linear relationship, validating the scaling law.
The linear fits yield the intercepts $\eta$ of -2.7(1)$\times$10$^{-4}$ and -3.2(3)$\times$10$^{-4}$ V$^{-1}$ cm, and the slopes $\xi$ of -8.2(2)$\times$10$^{-8}$ and -5.1(3)$\times$10$^{-8}$ V$^{-1}$ cm$^3$ $\Omega^2$ for the warming and cooling processes, respectively (\hyperref[sec:note9]{Supplemental Note~9}).
It indicates that the extrinsic mechanism (skew scattering) should dominate the nonlinear transport behaviors at high-temperature region.
Such a negligibly small intrinsic contribution is physically reasonable.
Previous transport and ARPES studies have established that EuTe$_4$ exhibits a fully gapped semiconductor behavior, with a maximum energy gap $E_g$ of approximately 196~meV at 55~K \cite{LvBQ2022}.
Since the intrinsic contribution (Berry curvature dipole) is governed by the integral of the Berry curvature derivatives over the states at the Fermi energy level, the absence of a Fermi surface in this CDW-gapped state naturally dictates that the Berry curvature dipole should be vanishingly small.
The finite contribution of the intrinsic mechanism could arise from thermally excited carriers at high temperature, as well as in-gap states induced by defects or domain boundaries in the extrinsic semiconductor regime, which typically inherit the character of their parent bands.
Furthermore, the linear fits reveal a noticeable difference in the slope $\xi$ between warming and cooling processes.
Since $\xi$ represents the skew scattering coefficient, it indicates that scattering process may change due to the different configurations of CDW domains between warming and cooling processes.

In contrast to the high-temperature regime, the low-temperature data deviate from the scaling law, particularly during the cooling process between 120 K and 220 K and the warming process between 180 K and 260 K [inset of Fig. 4(b)]. Furthermore, this deviation is accompanied by a sign change of $E_{abb}^{2\omega}/E_{b}^2$.
There are one possible explanation of the observed breakdown of scaling behavior and sign reversal behavior.
In the high-temperature regime ($>200$ K), the macroscopic domain evolution is primarily characterized by the transition between CDW and metallic states as well as the merging of domains with different polarization directions. This process concurrently modulates both of mobility and nonlinear Hall response, thus preserving the validity of the scaling behavior.
However, at low temperatures, the domain evolution mainly involves the realignment of polarization directions across different layers. While such 180$^\circ$ polarization flips leave the in-plane mobility largely unaffected, they profoundly alter the symmetry-dependent nonlinear Hall signal, naturally leading to a breakdown of the scaling behavior and the corresponding sign reversal.
Fig. 4(c) and \hyperref[sec:note10]{Supplemental Note~10} show the nonlinear responsivity $|E_{\perp}^{2\omega}|/(E_{\parallel}^{\omega})^2$ as a function of temperature for materials exhibiting the NLHE.
In most of reported materials, such as topological semimetals, large nonlinear responsivities only appear at low temperatures (below 30 K) and drop quickly as the temperature increases toward room temperature.
In contrast, bulk $\text{EuTe}_4$ is located in the high-temperature and high-responsivity region ($T\geq$ 100 K and $|E_{\perp}^{2\omega}|/(E_{\parallel}^{\omega})^2\geq$ 10$^{-1}$ $\mu$m V$^{-1}$).
In contrast, bulk $\text{EuTe}_4$ maintains a giant nonlinear responsivity across a wide temperature range up to 400 K. At room temperature, its nonlinear responsivity $|E_{\perp}^{2\omega}|/(E_{\parallel}^{\omega})^2$ reaches $\sim 3\text{--}6\ \mu\text{m}\ \text{V}^{-1}$ (and up to $\sim 15\ \mu\text{m}\ \text{V}^{-1}$ at 400 K), which is several orders of magnitude higher than most materials at room temperature, such as $\text{TaIrTe}_4$~\cite{KumarD}, few-layer $\text{WTe}_2$~\cite{MaQ,KangK}, and strained monolayer $\text{WSe}_2$~\cite{QinMS}.
More importantly, unlike other materials with obvious room-temperature NLHE, the nonlinear responsivity in $\text{EuTe}_4$ exhibits a pronounced thermal hysteresis loop between warming and cooling processes, providing an extra tunablity of NLHE.

EuTe$_{4}$ provides a promising platform to explore the novel properties and tunability of CDW-induced nonlinear electronic transport effects.
First, fabricating micro- or nano-scale devices could isolate single CDW domain, thereby circumventing the averaging effect and yielding even higher nonlinear signals that directly reflect the intrinsic CDW state.
Second, the CDW domains and their polarization directions could be tuned by strain, and the carrier density and the associated CDW phases in semiconducting EuTe$_{4}$ thin flakes can also be effectively tuned via electrostatic gating. Thus nonlinear transport effects may also be manipulated by these external fields effectively.
Combining this tunable giant nonlinear response with the pronounced thermal hysteresis origin to the CDW state, EuTe$_{4}$ demonstrates tremendous potential for advanced applications of NLHE, particularly in novel memory devices.

In summary, we have systematically investigated the nonlinear transport effects of bulk EuTe$_4$ crystals.
Our results reveal that the nonlinear Hall voltage is strongly coupled to the hysteretic CDW state, leading to a pronounced thermal memory effect that can modulate the signal by a factor of about 4.7 at 340 K.
While the high-temperature transport conforms to standard scaling law, the continuous dynamic reconfiguration of CDW domains at lower temperatures may lead to a breakdown of scaling behavior.
Despite this complex domain configuration, EuTe$_4$ sustains a large nonlinear Hall response up to 400 K, with room-temperature responsivities exceeding representative reported values for classic TaIrTe$_4$ and few-layer WTe$_2$ systems.
Our findings demonstrate that EuTe$_4$ is a highly promising platform for designing novel, tunable nonlinear electronic devices at room temperature.

\begin{acknowledgments}
This work was supported by the National Key R\&D Program of China (Grant Nos. 2023YFA1406500 and 2022YFA1403800), the National Natural Science Foundation of China (Grant Nos. 12625409, 12274459 and 52572288, and W2431011), the Shanghai Key Laboratory of Novel Extreme Condition Materials, China (Grant No. 22dz2260800), the Shanghai Science and Technology Committee, China (Grant No. 22JC1410300), the Fundamental Research Funds for the Central Universities, and the Research Funds of Renmin University of China (Grant No. 26XNKJ22).
\end{acknowledgments}

\noindent
$^*$Correspondence should be addressed to X. Z. (zhangxiaobupt@bupt.edu.cn) and H. C. L. (hlei@ruc.edu.cn).

\begingroup
\renewcommand{\label}[1]{}
\bibliographystyle{apsrev4-1}
\bibliography{references}
\endgroup

\clearpage
\onecolumngrid
\begin{bibunit}[apsrev4-1]
\makeatletter
\def\@extra@b@citeb{.si}
\def\@extra@binfo{.si}
\makeatother
\setcounter{equation}{0}
\setcounter{figure}{0}
\setcounter{table}{0}
\renewcommand{\theequation}{S\arabic{equation}}
\renewcommand{\thefigure}{S\arabic{figure}}
\renewcommand{\thetable}{S\arabic{table}}
\renewcommand{\theHequation}{supp.\arabic{equation}}
\renewcommand{\theHfigure}{supp.\arabic{figure}}
\renewcommand{\theHtable}{supp.\arabic{table}}

\pagenumbering{arabic}
\thispagestyle{empty}
\AddToShipoutPictureFG*{%
  \AtPageUpperLeft{%
    \setlength{\unitlength}{1bp}%
    \put(72,-81.963){\parbox[t]{468pt}{\centering
      \fontsize{14.4}{27}\selectfont\bfseries
      Supplementary Information for\\
      Thermally Hysteretic Room-Temperature Nonlinear Hall Effect\\
      in Bulk EuTe$_4$\par}}%
    \put(72,-169.136){\parbox[t]{468pt}{\centering
      \fontsize{12}{22}\selectfont
      \leftskip=0pt plus .3\hsize\relax\rightskip=\leftskip
      Yang Fu,$^{1,2}$ Xingyu Lv,$^{2}$ Shangjie Tian,$^{2}$ Ying Ma,$^{3}$ Xiao\\
      Zhang,$^{3,*}$ Cedomir Petrovic,$^{4,1,6}$ and Hechang Lei$^{2,5,*}$\par
      {\itshape
        $^{1}$Center for High Pressure Science and Technology\\
        Advanced Research (HPSTAR), Beijing 100193, China\\
        $^{2}$School of Physics and Key Laboratory of Quantum State\\
        Construction and Manipulation (Ministry of Education),\\
        Renmin University of China, Beijing 100872, China\\
        $^{3}$State Key Laboratory of Information Photonics and Optical\\
        Communications $\&$ School of Physical Science and Technology,\\
        Beijing University of Posts and Telecommunications, Beijing 100876, China\\
        $^{4}$Shanghai Key Laboratory of Material Frontiers\\
        Research in Extreme Environments (MFree),\\
        Shanghai Advanced Research in Physical Sciences (SHARPS), Shanghai 201203, China\\
        $^{5}$Xiongan Future Industrial Institute,\\
        Renmin University of China, Xiongan 071700, China\\
        $^{6}$Tsung-Dao Lee Institute, Shanghai Jiao Tong University, Shanghai 200240, China
        \hspace{\fontdimen2\font plus\fontdimen3\font minus\fontdimen4\font}\par}}}%
    \put(72,-707.491){\rule{131.04pt}{0.4pt}}%
    \put(72,-732.531){\fontsize{10}{14.5}\selectfont\spaceskip=2.443pt\relax
      $^*$Correspondence should be addressed to X. Z. (zhangxiaobupt@bupt.edu.cn) and H. C. L. (hlei@ruc.edu.cn).}%
    \put(296.349,-764.911){\fontsize{12}{21}\selectfont\thepage}%
  }%
}
\null
\newpage

\section*{Supplemental Note 1: Symmetry Analysis and Tensor Symmetries}
\label{sec:note1}

The room-temperature charge-density-wave (CDW) phase of bulk EuTe$_4$ belongs to the non-centrosymmetric space group $P2_1cn$ (No. 33) \cite{WuD}.
For macroscopic second-order nonlinear transport responses, the relevant symmetry is determined by the crystallographic point group.
In this context, microscopic translation-related symmetry operations, such as glide planes and screw axes, act effectively as simple mirror planes and rotation axes.
Consequently, the nonlinear transport properties of EuTe$_4$ are governed by the point group $C_{2v}$ ($mm2$),
which is identical to the point group of the well-known nonlinear Hall material WTe$_2$ and TaIrTe$_4$ \cite{SM_Ma2019,SM_Kumar2021}.

To derive the symmetry-allowed nonlinear conductivity tensor, we begin with the general phenomenological equation for the second-order nonlinear current response,

\begin{equation}
	J_i^{2\omega} = \sum_{j,k} \chi_{ijk} E_j^{\omega} E_k^{\omega}
\end{equation}

where the indices $i, j, k \in \{a, b, c\}$ denote the crystallographic directions.
Here, $J_i^{2\omega}$ is the $i$-th component of the second-order nonlinear current density, $E_j^{\omega}$ and $E_k^{\omega}$ are the components of the applied alternating electric field at the frequency $\omega$,
and $\chi_{ijk}$ represents the elements of the nonlinear conductivity tensor.
Because both the current density $\boldsymbol{J}^{2\omega}$ and the electric field $\boldsymbol{E}^{\omega}$ are polar vectors, the nonlinear conductivity $\chi_{ijk}$ transforms as a third-rank polar tensor.
Furthermore, since the electric fields commute ($E_j^{\omega} E_k^{\omega} = E_k^{\omega} E_j^{\omega}$), the tensor $\chi_{ijk}$ is symmetric with respect to its last two indices ($\chi_{ijk} = \chi_{ikj}$).
This allows us to express the tensor as a $3 \times 6$ matrix with 18 independent elements (Voigt notation),

\begin{equation}
	\begin{bmatrix} J_a^{2\omega}\\ J_b^{2\omega}\\ J_c^{2\omega}\end{bmatrix} = \begin{bmatrix}  \chi_{aaa} & \chi_{abb} & \chi_{acc} & \chi_{abc} & \chi_{aac} & \chi_{aab} \\  \chi_{baa} & \chi_{bbb} & \chi_{bcc} & \chi_{bbc} & \chi_{bac} & \chi_{bab} \\  \chi_{caa} & \chi_{cbb} & \chi_{ccc} & \chi_{cbc} & \chi_{cac} & \chi_{cab} \end{bmatrix} \begin{bmatrix} (E_a^{\omega})^2 \\ (E_b^{\omega})^2 \\ (E_c^{\omega})^2 \\ 2E_b^{\omega} E_c^{\omega} \\ 2E_a^{\omega} E_c^{\omega} \\ 2E_a^{\omega} E_b^{\omega} \end{bmatrix}
\end{equation}

Based on the crystal symmetry, the space group $P2_1cn$ possesses the following key symmetry operations:

\begin{enumerate}
	\item A glide plane perpendicular to the $b$-axis, which effectively acts as a mirror plane ($M_b$), transforming the coordinates as $(a, b, c) \to (a, -b, c)$.
	\item A glide plane perpendicular to the $c$-axis, which effectively acts as a mirror plane ($M_c$), transforming the coordinates as $(a, b, c) \to (a, b, -c)$.
	\item A two-fold screw axis along the $a$-axis, which effectively acts as a rotation axis ($C_{2a}$), transforming the coordinates as $(a, b, c) \to (a, -b, -c)$.
\end{enumerate}

The nonlinear conductivity tensor elements must remain invariant under these symmetry operations.
Because the operations involve only reflections or two-fold rotations along the principal axes, each spatial index is either preserved or inverted.
Consequently, a transformed tensor element $\chi_{i'j'k'}$ is equal to $\pm \chi_{ijk}$, where the negative sign applies if the operation inverts an odd number of its corresponding spatial coordinates.
To satisfy the invariance requirement ($\chi_{i'j'k'} = \chi_{ijk}$), any element that acquires a negative sign must be identically zero.
For example, under the $M_b$ mirror symmetry, any tensor element with an odd number of $b$ indices (such as $\chi_{aab}$) transforms to its negative ($\chi_{i'j'k'} = -\chi_{aab}$), and therefore must be zero.
Applying these parity rules to all 18 elements, we find that the only non-zero components are $\chi_{aaa}$, $\chi_{abb}$, $\chi_{acc}$, $\chi_{bab} = \chi_{bba}$, and $\chi_{cac} = \chi_{cca}$.
Thus, the macroscopic nonlinear transport equation can be elegantly written in the following matrix form for space group $P2_1cn$,

\begin{equation}
	\begin{bmatrix} J_a^{2\omega}\\ J_b^{2\omega}\\ J_c^{2\omega}\end{bmatrix} = \begin{bmatrix}  \chi_{aaa} & \chi_{abb} & \chi_{acc} & 0 & 0 & 0 \\  0 & 0 & 0 & 0 & 0 & \chi_{bba} \\  0 & 0 & 0 & 0 & \chi_{cca} & 0 \end{bmatrix} \begin{bmatrix} (E_a^{\omega})^2 \\ (E_b^{\omega})^2 \\ (E_c^{\omega})^2 \\ 2E_b^{\omega} E_c^{\omega} \\ 2E_a^{\omega} E_c^{\omega} \\ 2E_a^{\omega} E_b^{\omega} \end{bmatrix}
	\label{eq:tensor}
\end{equation}

When the driving electric field is confined within the $ab$-plane ($E_c^{\omega} = 0$), the formula further can be simplified as following,
\begin{equation}
	\begin{bmatrix} J_a^{2\omega}\\ J_b^{2\omega}\\ J_c^{2\omega}\end{bmatrix} = \begin{bmatrix}  \chi_{aaa} & \chi_{abb} & 0 & 0 & 0 & 0 \\  0 & 0 & 0 & 0 & 0 & \chi_{bba} \\  0 & 0 & 0 & 0 & 0 & 0 \end{bmatrix} \begin{bmatrix} (E_a^{\omega})^2 \\ (E_b^{\omega})^2 \\ 0 \\ 0 \\ 0 \\ 2E_a^{\omega} E_b^{\omega} \end{bmatrix}
\end{equation}

For a single Te layer, the relevant in-plane symmetry reduces to a single mirror plane. Under this mirror symmetry constraint, the two-dimensional nonlinear transport matrix is given by
\begin{equation}
	\begin{bmatrix} J_a^{2\omega}\\ J_b^{2\omega}\end{bmatrix}=\begin{bmatrix}  \chi_{aaa}&  \chi_{abb}&  0 \\  0&  0&  \chi_{bba} \end{bmatrix}\begin{bmatrix}(E_a^{\omega})^2 \\ (E_b^{\omega})^2 \\2E_a^{\omega}E_b^{\omega}\end{bmatrix}
\end{equation}

The macroscopic nonlinear effect is ultimately the superposition of contributions from all individual Te layers.

To generalize the response for an arbitrary in-plane measurement geometry, suppose the driving electric field $E^{\omega}$ is applied within the $ab$-plane at an angle $\theta$ relative to the $a$-axis.
The electric field components are $E_a^{\omega} = E^{\omega} \cos\theta$ and $E_b^{\omega} = E^{\omega} \sin\theta$.
Substituting these into the simplified matrix yields the nonlinear current components along the principal axes,
\begin{equation}
	\begin{aligned}
		J_a^{2\omega} &= (E^{\omega})^2 (\chi_{aaa} \cos^2\theta + \chi_{abb} \sin^2\theta) \\
		J_b^{2\omega} &= 2 (E^{\omega})^2 \chi_{bba} \sin\theta \cos\theta
	\end{aligned}
\end{equation}

The corresponding longitudinal ($J_{\parallel}^{2\omega}$) and transverse ($J_{\perp}^{2\omega}$) nonlinear currents measured relative to the applied field direction can be obtained through coordinate rotation,

\begin{equation}
	\begin{aligned}
		J_{\parallel}^{2\omega} &= J_a^{2\omega} \cos\theta + J_b^{2\omega} \sin\theta = (E^{\omega})^2 \cos\theta \left[ \chi_{aaa} \cos^2\theta + (\chi_{abb} + 2\chi_{bba}) \sin^2\theta \right] \\
		J_{\perp}^{2\omega} &= -J_a^{2\omega} \sin\theta + J_b^{2\omega} \cos\theta = (E^{\omega})^2 \sin\theta \left[ (2\chi_{bba} - \chi_{aaa}) \cos^2\theta - \chi_{abb} \sin^2\theta \right]
	\end{aligned}
\end{equation}
where $J_{\parallel}^{2\omega}$ and $J_{\perp}^{2\omega}$ are the nonlinear current densities parallel and perpendicular to the applied electric field $E^{\omega}$, respectively,
$\theta$ represents the angle between the electric field and the crystallographic $a$-axis, and $\chi_{aaa}$, $\chi_{abb}$,
and $\chi_{bba}$ are the independent non-zero tensor elements of the nonlinear conductivity dictated by the $P2_1cn$ symmetry.

Crucially, these equations demonstrate that any slight misalignment of the electrical contacts relative to the principal crystallographic axes will induce residual symmetry-forbidden signals.
For instance, if the applied electric field is nominally directed along the $a$-axis but suffers from a small misalignment angle $\theta$,
the transverse nonlinear current $J_{\perp}^{2\omega}$ (which corresponds to the symmetry-forbidden $V_{baa}^{2\omega}$ signal in the main text) is no longer zero,
but becomes linearly proportional to the misalignment,

\begin{equation}
	J_{\perp}^{2\omega}(\theta \to 0) \approx (E^{\omega})^2 \theta (2\chi_{bba} - \chi_{aaa})
\end{equation}

Similarly, if the electric field is nominally applied along the $b$-axis (corresponding to $\theta \approx \pi/2$), we can define a small deviation angle $\delta = \pi/2 - \theta$.
The longitudinal nonlinear current $J_{\parallel}^{2\omega}$ (which corresponds to the symmetry-forbidden $V_{bbb}^{2\omega}$ signal and is exactly zero when perfectly aligned) emerges as

\begin{equation}
	J_{\parallel}^{2\omega}(\delta \to 0) \approx (E^{\omega})^2 \delta (\chi_{abb} + 2\chi_{bba})
\end{equation}

Therefore, the inevitable small deviations in device fabrication or crystal cleavage naturally account for the finite residual symmetry-forbidden nonlinear voltages observed in the transport measurements.

\newpage
\section*{Supplemental Note 2: Calculation of Nonlinear Conductivities}
\label{sec:note2}

In transport measurements, the second-harmonic signals are detected under near open-circuit conditions. Under these conditions,
the intrinsic nonlinear current density $\boldsymbol{J}^{2\omega}$ generates a corresponding nonlinear electric field $\boldsymbol{E}^{2\omega}$ to maintain zero net current, which satisfies the Ohm's law,

\begin{equation}
	\boldsymbol{E}^{2\omega} = \boldsymbol{\rho} \cdot \boldsymbol{J}^{2\omega}
\end{equation}

where $\boldsymbol{\rho}$ is the linear resistivity tensor for anisotropic EuTe$_4$,

\begin{equation}
	\boldsymbol{\rho} = \begin{pmatrix} \rho_a & 0 & 0 \\ 0 & \rho_b & 0 \\ 0 & 0 & \rho_c \end{pmatrix}
\end{equation}

Thus, the components of the intrinsic nonlinear current density can be deduced as $J_i^{2\omega} = E_i^{2\omega} / \rho_i$ (for $i = a, b, c$).
Both the applied primary electric field and the generated nonlinear electric field intensities are obtained by dividing their corresponding measured voltages by the distance between the respective measurement electrodes.

To experimentally determine the nonlinear conductivities, we express the applied primary electric field $E^{\omega}$ using the applied alternating current $I^{\omega}$,
the distance between the longitudinal voltage terminals $l$, the sample width $w$, and the sample thickness $d$.
When the primary current is applied along the $b$-axis, the electric field is $E_b^{\omega} = \rho_b j_b^{\omega} = I_b^{\omega} \rho_b / (wd)$.
Using the transverse nonlinear current $J_a^{2\omega} = E_a^{2\omega}/\rho_a$ and the transverse nonlinear electric field $E_a^{2\omega} = V_{abb}^{2\omega}/w$,
and assuming $\rho_a\approx\rho_b$, the transverse (nonlinear Hall) conductivity $\chi_{abb} = J_a^{2\omega}/(E_b^{\omega})^2$ can be expressed as
\begin{equation}
	\chi_{abb} = \frac{V_{abb}^{2\omega} w d^2}{(I_b^{\omega})^2 \rho_b^2 \rho_a} \approx \frac{V_{abb}^{2\omega} w d^2}{(I_b^{\omega})^2 \rho_b^3}
\end{equation}

Similarly, when the primary current is applied along the $a$-axis, the electric field is $E_a^{\omega} = \rho_a j_a^{\omega} = I_a^{\omega} \rho_a / (wd)$.
Using the longitudinal nonlinear current $J_a^{2\omega} = E_a^{2\omega}/\rho_a$ and $E_a^{2\omega} = V_{aaa}^{2\omega}/l$,
the longitudinal nonlinear conductivity $\chi_{aaa} = J_a^{2\omega}/(E_a^{\omega})^2$ can be expressed as

\begin{equation}
	\chi_{aaa} = \frac{V_{aaa}^{2\omega} w^2 d^2}{(I_a^{\omega})^2 \rho_a^3 l}
\end{equation}

\newpage
\section*{Supplemental Note 3: Crystal Growth and Experimental Methods}
\label{sec:note3}
EuTe$_4$ single crystals are grown using the self-flux method. Eu chunks (99.9 \%) and Te chunks (99.999 \%) were mixed in a molar ratio of 1 : 15 and loaded into a quartz crucible,
which was then sealed in a quartz ampoule under high vacuum (10$^{-2}$ Pa).
The sealed ampoule was heated to 1123 K over 24h, maintained there for 48h, then cooled at a rate of 4 K/h to 693 K.
After maintaining this temperature for one week, the mixture was centrifuged to remove flux and the plate-like EuTe$_4$ single crystals can be obtained.

The devices measured along the two different current directions along the crystallographic axes (as discussed in the main text) were cut from the same one single crystal.
The crystallographic axes of the crystals were accurately determined using a X-ray diffractometer equipped with a goniometer (Bruker D8 Advance).
After carefully cleaning the crystal surfaces, platinum (Pt) electrical contacts were deposited via magnetron sputtering through a hard mask.
Pt wires were then attached using conductive epoxy and baked inside a glove box to form robust and clean electrical connections.

For the transport measurements, an AC current was applied using a Keithley 6221 current source, and the corresponding primary and harmonic voltages were recorded using two NF LI5650 lock-in amplifiers.
When measuring the contact resistance ($R_{\rm c}$), one LI5650 lock-in amplifier was utilized as the AC voltage source.
The applied current was simultaneously monitored by measuring the voltage drop across a 1 k$\Omega$ series resistor using the other lock-in amplifier.
The $R_{\rm c}$ was subsequently derived from the ratio of the output voltage of the LI5650 to the measured current.
All temperature-dependent transport measurements were conducted in a home-built high-vacuum cryogenic system, enabling precise temperature control over a wide range from 3 K up to 400 K.

\newpage
\section*{Supplemental Note 4: Frequency Dependence of the Nonlinear Signal}
\label{sec:note4}
\begin{figure}[H]
	\centering\includegraphics[width=1\textwidth]{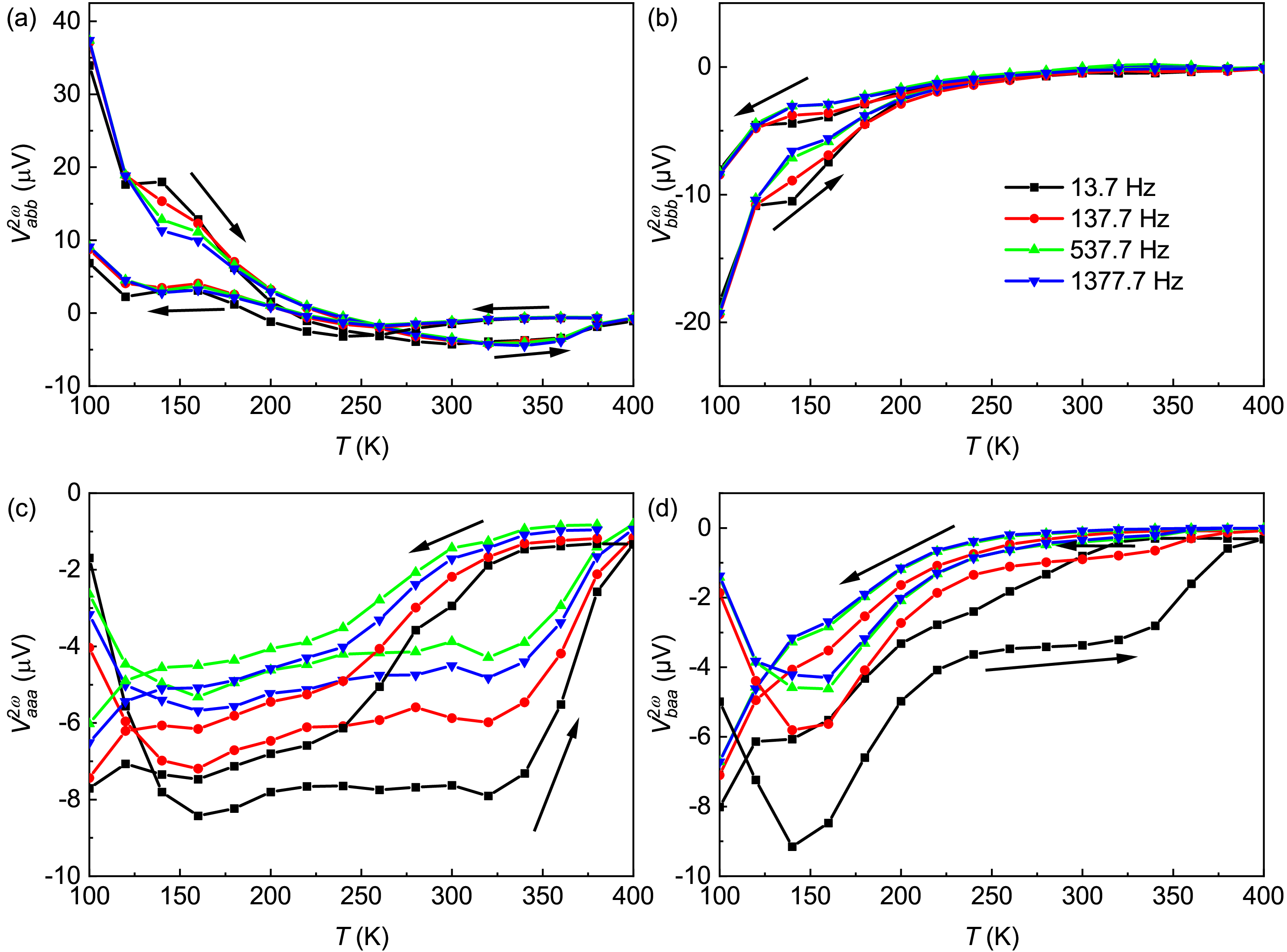}
	\caption{Frequency dependence of the second-harmonic voltages. Temperature-dependent (a) $V_{abb}^{2\omega}$, (b) $V_{bbb}^{2\omega}$, (c) $V_{aaa}^{2\omega}$, and (d) $V_{baa}^{2\omega}$ under various frequencies.
		The black arrows indicate the temperature sweep directions.
		The inset in (c) shows the enlarged view of $V_{aaa}^{2\omega}$ at the high-temperature region.}
	\label{fig:freq}
\end{figure}

To ensure that the observed nonlinear signals originate from the intrinsic physical properties of EuTe$_4$ rather than parasitic artifacts, we investigated the frequency dependence of the second-harmonic responses.
As shown in Fig. \ref{fig:freq}, the $V_{abb}^{2\omega}$ exhibits a slight frequency dependence, but its overall behavior remains highly consistent.
This indicates that the intrinsic nonlinear signal dominates the voltage response, thereby largely ruling out extrinsic contributions such as parasitic capacitive coupling and local Joule heating.
In contrast, the $V_{aaa}^{2\omega}$ displays a relatively stronger frequency dependence compared to the $V_{abb}^{2\omega}$.
This difference arises because the longitudinal measurement is more susceptible to parasitic capacitance and thermal gradient effects.
Specifically, Joule heating ($\propto (I^\omega)^2$) primarily generates longitudinal thermal gradients, which produce a significant $2\omega$ thermoelectric voltage along the primary current path.
Consequently, the $V_{abb}^{2\omega}$ is largely unaffected by these longitudinal thermal gradients and serves as a more reliable probe for the nonlinear Hall effect.
Nevertheless, despite this moderate frequency dependence in the longitudinal channel, the overall temperature-dependent behavior remains consistent across the measured frequency range.

\newpage
\section*{Supplemental Note 5: High-Temperature Nonlinear Hall Effect}
\label{sec:note5}

\begin{figure}[H]
\centering\includegraphics[width=0.6\textwidth]{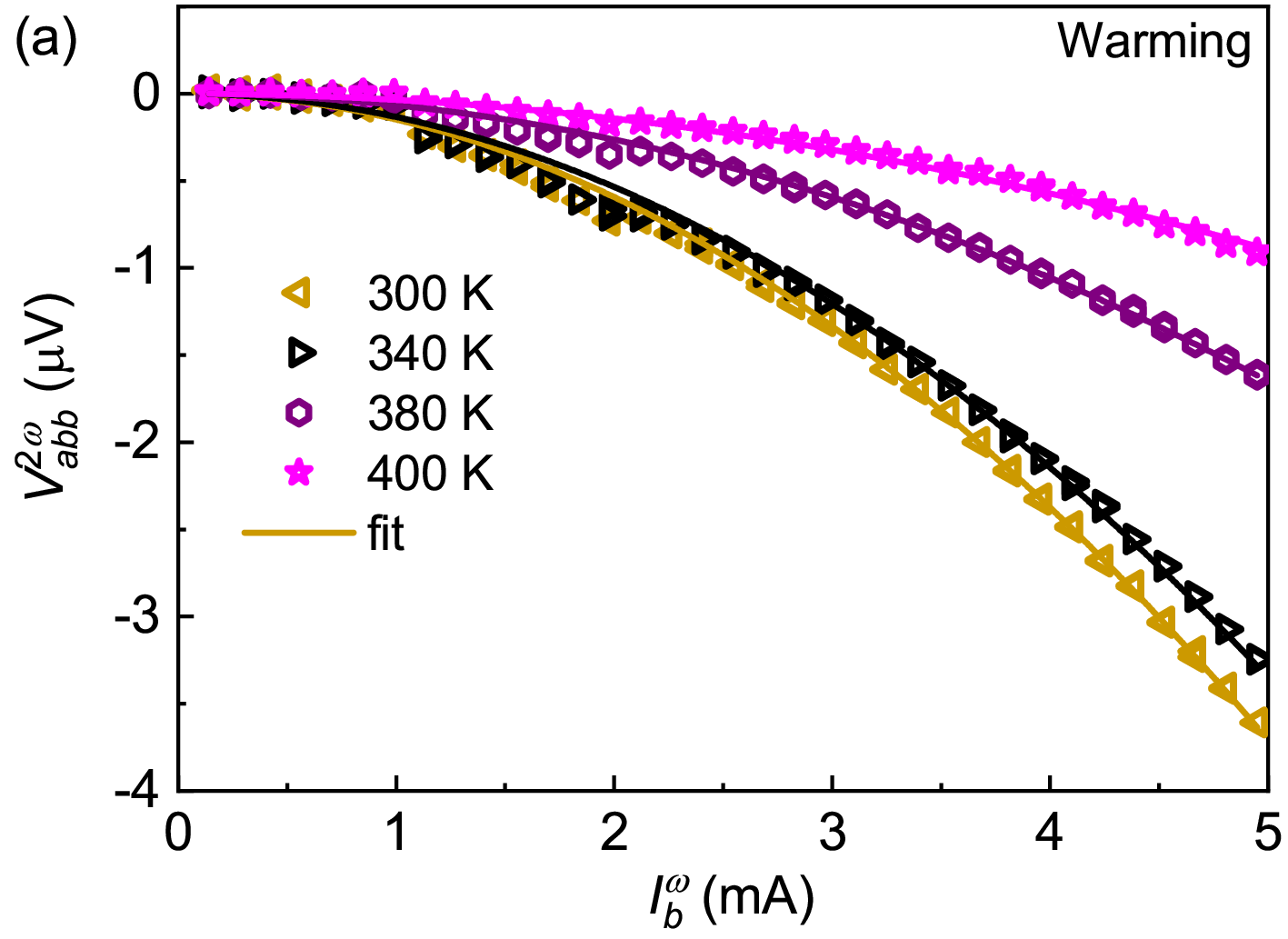}
\caption{Nonlinear Hall effect between 300 K--400 K. (a) $V_{abb}^{2\omega}$ as a function of $I_b^{\omega}$ during the warming process from 300 K to 400 K. The solid lines represent the quadratic fits.}
\label{fig:hightemp}
\end{figure}

As discussed in the main text, the nonlinear transport response in EuTe$_4$ exhibits a robust quadratic dependence on the applied current $I^{\omega}$ across the high-temperature range.
Figure \ref{fig:hightemp} displays the detailed $V_{abb}^{2\omega}$ curves measured during the warming process from 300 K to 400 K,
confirming that the nonlinear signal is well fitted by a quadratic relationship within this high-temperature regime.

\newpage
\section*{Supplemental Note 6: Nonlinear Hall Effect and Contact Resistance at Low Temperatures}
\label{sec:note6}

\begin{figure}[H]
\centering\includegraphics[width=1\textwidth]{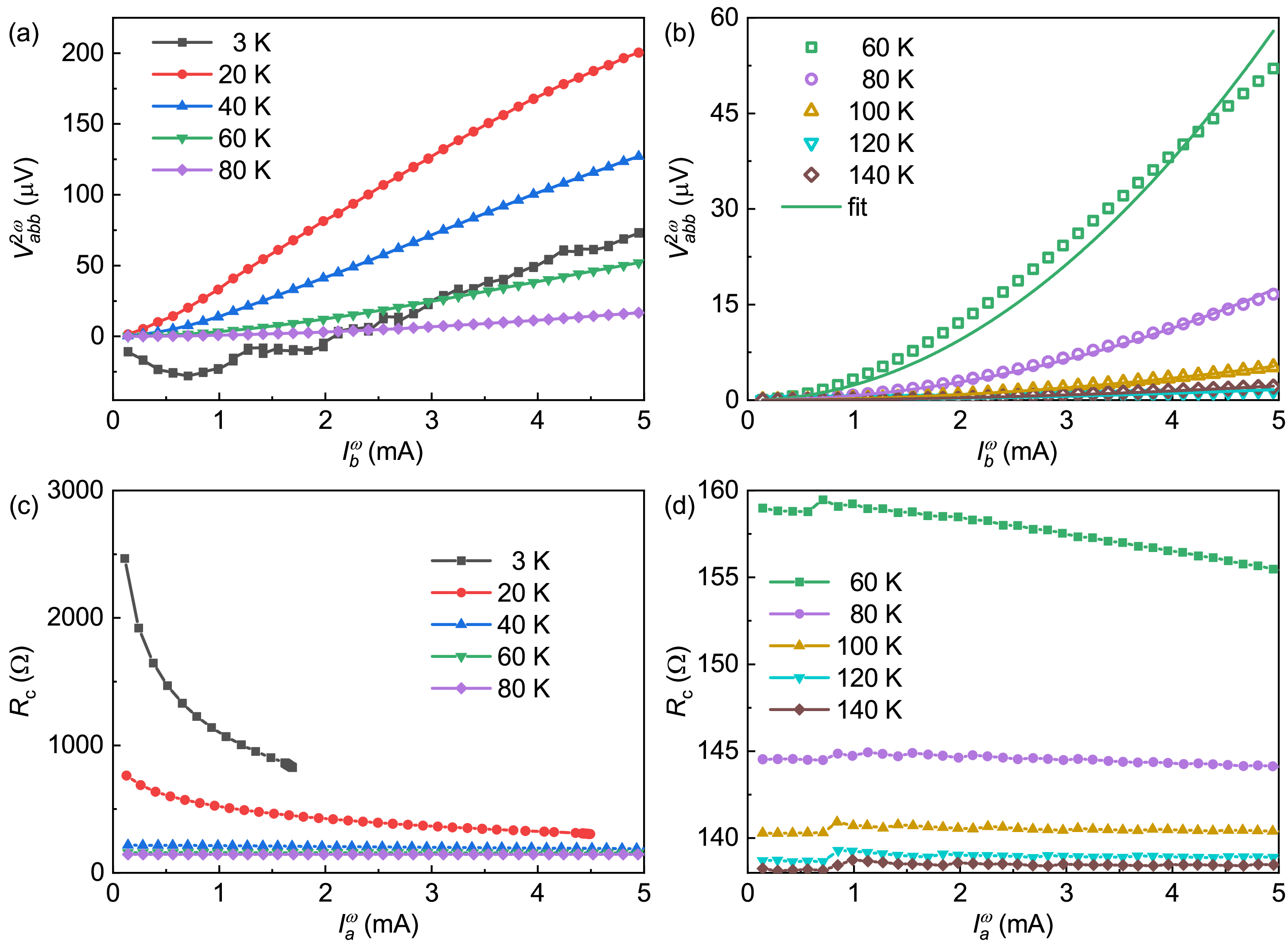}
\caption{Nonlinear Hall effect and contact resistance at low temperatures.
Second-harmonic voltage $V_{abb}^{2\omega}$ versus the applied current $I_{b}^{\omega}$ measured during the warming process at (a) 3 K to 80 K and (b) 60 K to 140 K.
The solid lines in (b) are quadratic fits, showing a clear deviation from the expected quadratic behavior below 100 K.
Current dependence of $R_{\rm c}$ at (c) 3 K to 80 K and (d) 60 K to 140 K.
At low temperatures, $R_{\rm c}$ increases drastically and exhibits a pronounced nonlinear dependence on the applied current,
indicative of the formation of Schottky barriers, whereas it is small and essentially current-independent at higher temperatures.}
\label{fig:contact}
\end{figure}

As discussed in the main text, while the second-harmonic voltage strictly adheres to the expected quadratic dependence on the applied current ($I^{\omega}$) at high temperatures,
this relationship begins to break down upon cooling below 100 K.
As shown in Figs. \ref{fig:contact}(a) and \ref{fig:contact}(b),
the nonlinear Hall voltages measured during the warming process exhibit a clear deviation from the typical quadratic behavior at low temperatures.
To elucidate the origin of this deviation, we examined the current dependence of $R_{\rm c}$ at various temperatures, as shown in Figs. \ref{fig:contact}(c) and \ref{fig:contact}(d).
Upon cooling below 40 K, the $R_{\rm c}$ increases drastically, reaching several kilo-ohms.
More importantly, it exhibits a pronounced current dependence, decreasing sharply as the current increases.
This strong current dependence is a characteristic signature of the formation of Schottky barriers at the interface between the metallic electrodes and the EuTe$_4$ crystal.
In contrast, at higher temperatures ($\ge 100$ K) [Fig. \ref{fig:contact}(d)], the $R_{\rm c}$ is significantly lower (around 150 $\Omega$)
and becomes basically independent of the applied current, confirming the presence of good Ohmic contacts.
The emergence of Schottky barriers at low temperatures could induce parasitic capacitive effects and non-ideal transport behaviors,
which contaminate the intrinsic bulk nonlinear Hall response and cause the deviation from the quadratic scaling law.
Therefore, to ensure the reliability of our analysis, the scaling law discussions in the main text are strictly confined to the high-temperature regime ($\ge 100$ K) with Ohmic contacts.

\newpage
\section*{Supplemental Note 7: Nonlinear Hall Effect During the Cooling Process}
\label{sec:note7}

\begin{figure}[H]
\centering\includegraphics[width=0.8\textwidth]{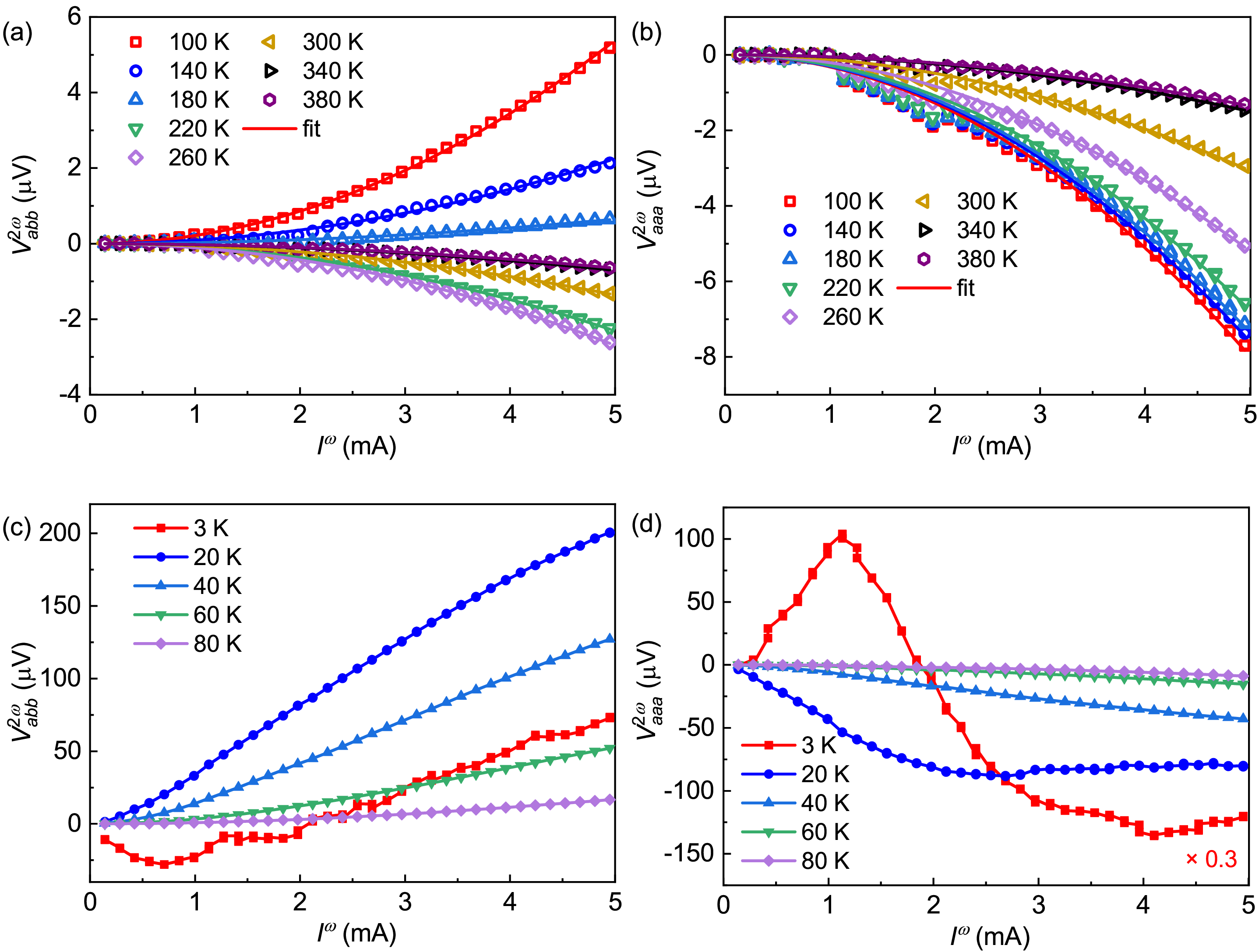}
\caption{Nonlinear Hall effect during the cooling process.
Second-harmonic voltages (a) $V_{abb}^{2\omega}$ and (b) $V_{aaa}^{2\omega}$ as a function of the applied current $I^{\omega}$ from 100 K to 380 K.
The solid lines in (a) and (b) represent quadratic fits. (c) $V_{abb}^{2\omega}$ and (d) $V_{aaa}^{2\omega}$ as a function of $I^{\omega}$ from 3 K to 80 K.
In (d), the data for $V_{aaa}^{2\omega}$ at 3 K are multiplied by 0.3 for clarity.
The nonlinear signals exhibit a typical quadratic behavior at high temperatures and deviate at low temperatures.}
\label{fig:cooling}
\end{figure}

To further investigate the nonlinear transport behaviors associated with the CDW transition, we also measured the second-harmonic voltages during the cooling process (from 400 K down to 3 K).
As shown in Fig. \ref{fig:cooling}, the nonlinear Hall effect during the cooling process exhibits a similar quadratic current dependence at high temperatures.
Furthermore, upon cooling below 100 K, the nonlinear response again deviates from the typical quadratic scaling,
consistent with the emergence of Schottky barriers and the increase of contact resistance at low temperatures.

\newpage
\section*{Supplemental Note 8: Temperature Dependence of Longitudinal Nonlinear Conductivity \texorpdfstring{$\chi_{aaa}$}{chi(aaa)}}
\label{sec:note8}

\begin{figure}[H]
	\centering\includegraphics[width=0.6\textwidth]{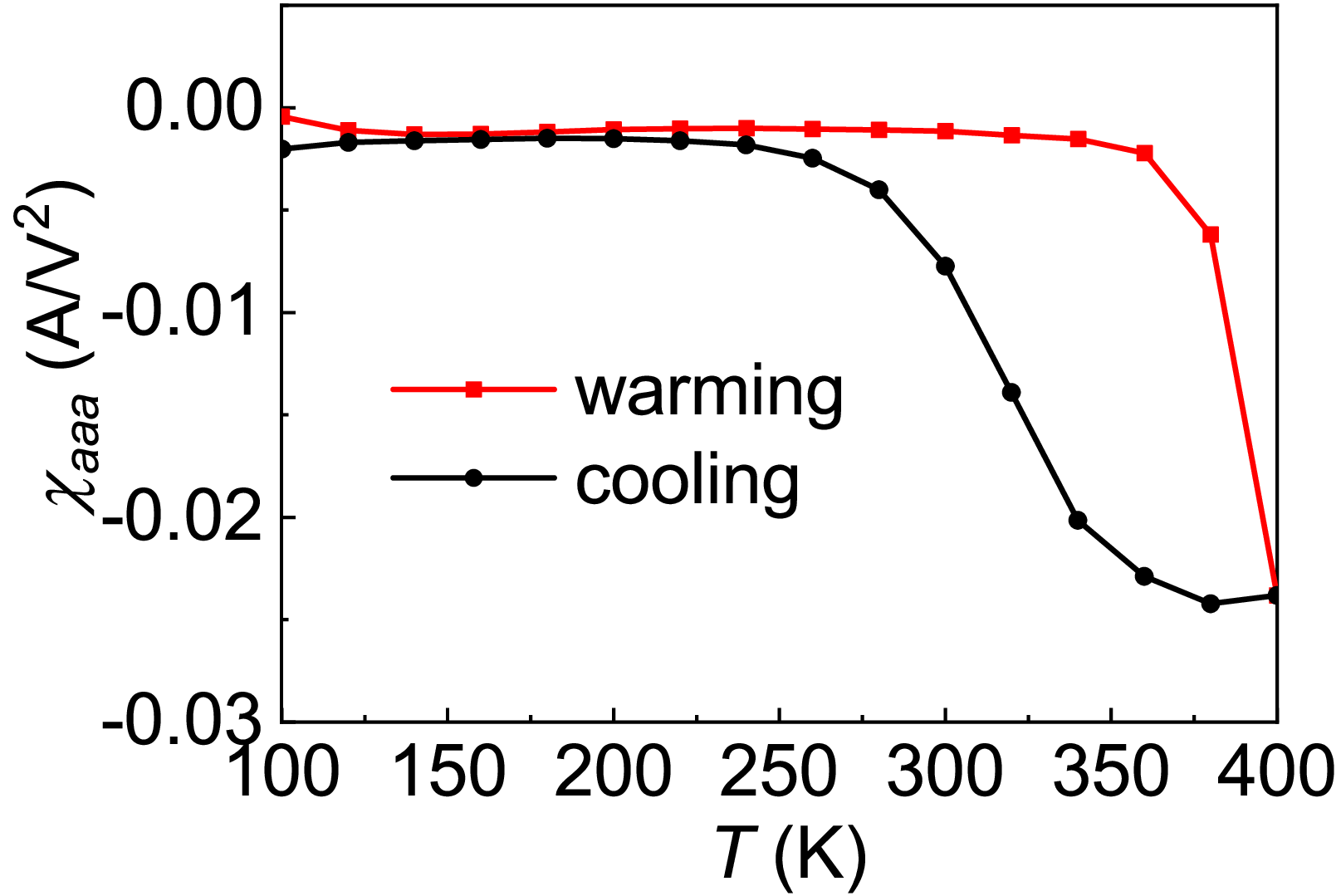}
	\caption{Temperature dependence of $\chi_{aaa}(T)$ measured during the warming and cooling processes between 100 K and 400 K.}
	\label{fig:SM5}
\end{figure}

\newpage
\section*{Supplemental Note 9: Detailed Fitting Curves for Scaling Law}
\label{sec:note9}

\begin{figure}[H]
	\centering\includegraphics[width=0.6\textwidth]{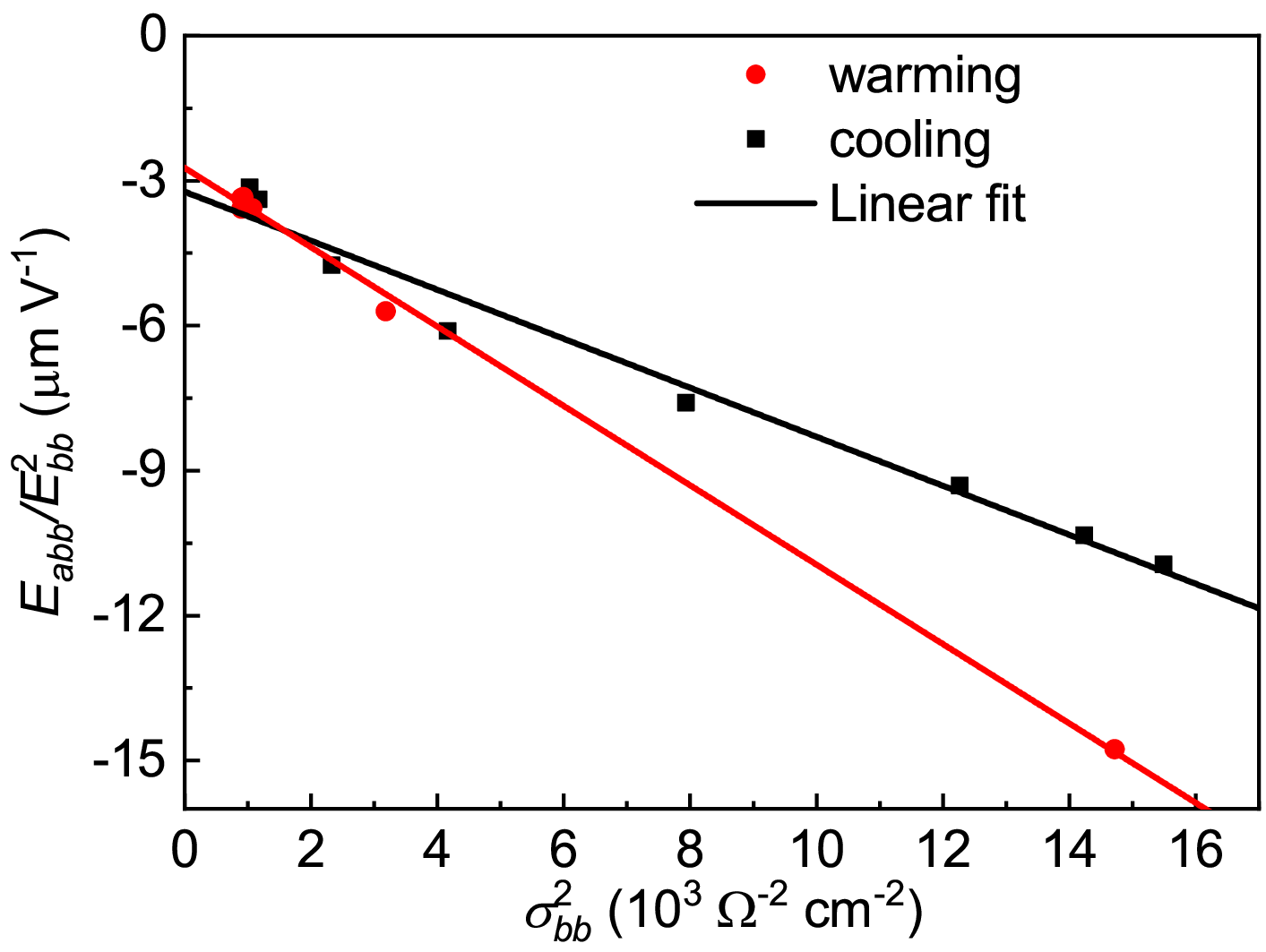}
	\caption{Linear fits of the scaling law $E_{abb}/E_{b}^2$ versus $\sigma_{bb}^2$ at high temperatures.
		The red circles and black squares represent the experimental data for the warming (280 K--400 K) and cooling (240 K--380 K) branches, respectively.
		The solid lines denote the corresponding linear fits.}
	\label{fig:scaling}
\end{figure}

\newpage
\section*{Supplemental Note 10: Comparison of Nonlinear Transport Parameters}
\label{sec:note10}

\begingroup
\color{black}
\renewcommand{\baselinestretch}{1}
\fontsize{9}{12}\selectfont
\DefTblrTemplate{middlehead,lasthead}{default}{Table \thetable\ (continued)}
\DefTblrTemplate{contfoot}{default}{}
\DefTblrTemplate{lastfoot}{default}{%
  \parbox{\linewidth}{\fontsize{9}{11}\selectfont
Notes: 1, directly reported in text or tables; 2, extracted from figures or calculated using reported parameters and stated dimensions; 3, calculated using dimensions estimated from micrographs or scale bars.}
}
\begin{longtblr}[
  caption = {Nonlinear Hall responsivity of EuTe$_4$ and typical materials reported in the literature.},
  label = {tab:comparison},
]{
  width = \textwidth,
  colspec = {Q[l, wd=3.12cm] Q[c, wd=1.3cm] Q[c, wd=4.0cm] X[l] Q[c, wd=1.65cm]},
  rowhead = 1,
  cells = {valign=t, fg=black, font=\fontsize{9}{12}\selectfont},
  row{1} = {valign=m},
  colsep = 3pt,
  rowsep = 2pt,
  column{1} = {leftsep=0pt},
  column{5} = {rightsep=0pt},
  hline{1,2,Z} = {0.5pt},
}
\textbf{Material} & \boldmath$T$\ (K) & \boldmath$E_{\perp}^{2\omega}/(E_{\parallel}^{\omega})^2$\ ($\mu$m\,V$^{-1}$) & \textbf{Remarks} & \textbf{Ref.} \\
EuTe$_4$ & 400 & $14.76$ & & \mbox{This work} \\*
EuTe$_4$ & 300 & $3.581$ & Warming branch & \mbox{This work} \\*
EuTe$_4$ & 300 & $6.108$ & Cooling branch & \mbox{This work} \\*
EuTe$_4$ & 100 & $54.1$ & Warming branch & \mbox{This work} \\*
EuTe$_4$ & 100 & $11.15$ & Cooling branch & \mbox{This work} \\
WTe$_2$\textsuperscript{\,3} & 10 & $15$ & 2 layers & \cite{SM_Ma2019} \\
WTe$_2$\textsuperscript{\,2} & 10 & $0.00141$ & 5 layers & \cite{SM_Kang2019} \\
BP/MnBi$_2$Te$_4$/BP\textsuperscript{\,3} & 1.8 & $2.3$ & 6 SL; Gated & \cite{Gao2023} \\*
BP/MnBi$_2$Te$_4$/BP\textsuperscript{\,3} & 11 & $0.8$ & 6 SL & \cite{Gao2023} \\
MnBi$_2$Te$_4$\textsuperscript{\,2} & 2 & $2.7$ & 4 SL & \cite{Wang2023} \\*
MnBi$_2$Te$_4$\textsuperscript{\,2} & 2 & $4.8$ & 6 SL & \cite{Wang2023} \\
TaIrTe$_4$\textsuperscript{\,2} & 300 & $0.0047$ & $t = 16$~nm & \cite{SM_Kumar2021} \\*
TaIrTe$_4$\textsuperscript{\,2} & 2 & $0.0248$ & $t = 16$~nm & \cite{SM_Kumar2021} \\
Bi$_2$Se$_3$\textsuperscript{\,2} & 10 & $0.00115$ & 20 QL & \cite{SM_He2021} \\*
Bi$_2$Se$_3$\textsuperscript{\,2} & 200 & $3.9\times10^{-4}$ & 20 QL & \cite{SM_He2021} \\
$T_d$-MoTe$_2$\textsuperscript{\,1} & 2 & $4.3\times10^{5}$ & $t = 127$ nm; $c$-axis output & \cite{SM_Tiwari2021} \\
Ce$_3$Bi$_4$Pd$_3$\textsuperscript{\,1} & $\approx1.7$ & $3\times10^{3}$ &  & \cite{SM_Dzsaber2021} \\
WSe$_2$\textsuperscript{\,3} & 1.5 & $1\times10^{4}$ & Twisted bilayer; Gated & \cite{Huang2023WSe2} \\
Graphene\textsuperscript{\,2} & 1.7 & $5.1\times10^{3}$ & Twisted bilayer; Gated  & \cite{Duan2022Graphene} \\*
Graphene\textsuperscript{\,2} & 80 & $4.6$ & Twisted bilayer; Gated & \cite{Duan2022Graphene} \\
WSe$_2$\textsuperscript{\,2} & 140 & $0.028$ & Strained monolayer; Gated & \cite{SM_Qin2021} \\
WTe$_2$\textsuperscript{\,2} & 5 & $91$ & 12 layers; in-plane d.c. electric field & \cite{Ye2023WTe2BCD} \\
BaMnSb$_2$\textsuperscript{\,2} & 300 & $170$ &  & \cite{Min2023BaMnSb2} \\*
BaMnSb$_2$\textsuperscript{\,2} & 206 & $1.3\times10^{3}$ &  & \cite{Min2023BaMnSb2} \\
LaAlO$_3$/SrTiO$_3$\textsuperscript{\,2} & 1.5 & $2.6$ & (111) interface; Gated & \cite{Lesne2023} \\
Bi\textsuperscript{\,2} & 300 & $0.013$ & $t=100$nm & \cite{Makushko2024Bi} \\*
Bi\textsuperscript{\,2} & 333 & $0.019$ & $t=100$nm & \cite{Makushko2024Bi} \\
Te\textsuperscript{\,2} & 300 & $0.075$ & Gated & \cite{Cheng2024Te} \\*
Te\textsuperscript{\,2} & 300 & $0.015$ &  & \cite{Cheng2024Te} \\*
Te\textsuperscript{\,2} & 200 & $0.075$ &  & \cite{Cheng2024Te} \\
PbTaSe$_2$\textsuperscript{\,2} & 300 & $0.0118$ & $t=123$nm & \cite{Itahashi2022PbTaSe2} \\*
PbTaSe$_2$\textsuperscript{\,2} & 2.8 & $1.45\times10^{4}$ & $t=123$nm; superconducting mixed state & \cite{Itahashi2022PbTaSe2} \\*
PbTaSe$_2$\textsuperscript{\,2} & 6.5 & $70$ & $t=123$nm; normal state & \cite{Itahashi2022PbTaSe2} \\
ZrTe$_5$\textsuperscript{\,3} & 2 & $0.75$ & $t=60$nm & \cite{Wang2024ZrTe5} \\
BiTeBr\textsuperscript{\,1} & 300 & $0.57$ & $t=4$~nm & \cite{Lu2024BiTeBr} \\*
BiTeBr\textsuperscript{\,2} & 20 & $7.2$ & $t=4$~nm & \cite{Lu2024BiTeBr} \\*
BiTeBr\textsuperscript{\,2} & 350 & $0.42$ & $t=4$~nm & \cite{Lu2024BiTeBr} \\
SnTe\textsuperscript{\,2} & 300 & $0.0415$ & $t=20$nm; (001) plane & \cite{Hu2025SnTe} \\*
SnTe\textsuperscript{\,2} & 10 & $0.022$ & $t=20$nm; (001) plane & \cite{Hu2025SnTe} \\
WTe$_2$\textsuperscript{\,2} & 2 & $44$ & $t=35$nm & \cite{Li2023WTe2Current} \\
Mn(Bi$_{0.7}$Sb$_{0.3}$)$_4$Te$_7$\textsuperscript{\,3} & 4 & $0.215$ & $t=20$nm & \cite{Wang2024MnBi4Te7} \\*
Mn(Bi$_{0.7}$Sb$_{0.3}$)$_4$Te$_7$\textsuperscript{\,3} & 200 & $0.015$ & $t=20$nm & \cite{Wang2024MnBi4Te7} \\
MnBi$_2$Te$_4$\textsuperscript{\,2} & 30 & $3\times10^{-4}$ & 7 SL & \cite{Liu2024MnBi2Te4} \\*
MnBi$_2$Te$_4$\textsuperscript{\,2} & 300 & $1.1\times10^{-4}$ & 7 SL & \cite{Liu2024MnBi2Te4} \\*
MnBi$_2$Te$_4$\textsuperscript{\,2} & 315 & $1.06\times10^{-4}$ & 7 SL & \cite{Liu2024MnBi2Te4} \\*
MnBi$_2$Te$_4$\textsuperscript{\,2} & 10 & $0.028$ & Gated & \cite{Liu2024MnBi2Te4} \\*
MnBi$_2$Te$_4$\textsuperscript{\,2} & 325 & $6.2\times10^{-5}$ & 14 SL & \cite{Liu2024MnBi2Te4} \\
Te\textsuperscript{\,2} & 4 & $3\times10^{3}$ & Gated & \cite{Kim2024TeGate} \\*
Te\textsuperscript{\,2} & 20 & $5.3\times10^{3}$ & 28 nm & \cite{Kim2024TeGate} \\
Cd$_3$As$_2$\textsuperscript{\,2} & 270 & $0.13$ & $t=95$ nm; at zero field; Gated & \cite{Zhao2025Cd3As2QMD} \\
TaIrTe$_4$\textsuperscript{\,2} & 20 & $1.1$ & $t=8$ nm; at zero field & \cite{Wang2024TaIrTe4Orbital} \\*
TaIrTe$_4$\textsuperscript{\,2} & 250 & $0.015$ & $t=8$ nm; at zero field & \cite{Wang2024TaIrTe4Orbital} \\
WSe$_2$\textsuperscript{\,2} & 2 & $31$ & Twisted bilayer; Gated & \cite{Cao2025WSe2Scaling} \\*
WSe$_2$\textsuperscript{\,2} & 23 & $10.2$ & Twisted bilayer; Gated & \cite{Cao2025WSe2Scaling} \\
6R-TaS$_2$\textsuperscript{\,2} & 300 & $0.11$ &  & \cite{Wu2025TaS2} \\
Fe$_5$GeTe$_2$/Al$_x$O$_3$\textsuperscript{\,2} & 5 & $2.93$ &  & \cite{Zhong2024Fe5GeTe2} \\*
Fe$_5$GeTe$_2$/Al$_x$O$_3$\textsuperscript{\,2} & 80 & $0.67$ &  & \cite{Zhong2024Fe5GeTe2} \\
TaIrTe$_4$\textsuperscript{\,1} & 4 & $172$ & $t=88.2$ nm; in-plane d.c. electric field & \cite{Yang2025TaIrTe4Control} \\
$T_d$-WTe$_2$\textsuperscript{\,1} & 10 & $1.1$ & 3 layers & \cite{Wu2025WTe2Growth} \\
CdGeAs$_2$\textsuperscript{\,2} & 300 & $85$ &  & \cite{Lee2026CdGeAs2} \\*
CdGeAs$_2$\textsuperscript{\,2} & 316 & $100$ &  & \cite{Lee2026CdGeAs2} \\*
CdGeAs$_2$\textsuperscript{\,2} & 350 & $79$ &  & \cite{Lee2026CdGeAs2} \\
Bi$_2$Te$_3$\textsuperscript{\,1} & 300 & $0.006$ & 30 nm & \cite{Shen2025Bi2Te3} \\*
Bi$_2$Te$_3$\textsuperscript{\,2} & 300 & $0.0064$ & 50 nm & \cite{Shen2025Bi2Te3} \\*
Bi$_2$Te$_3$\textsuperscript{\,1} & 200 & $0.012$ & 30 nm & \cite{Shen2025Bi2Te3} \\*
Bi$_2$Te$_3$\textsuperscript{\,2} & 2 & $0.04$ & 30 nm & \cite{Shen2025Bi2Te3} \\
MoS$_2$\textsuperscript{\,1} & 10 & $20$ &  & \cite{Zhou2026MoS2} \\
Graphene\textsuperscript{\,2} & 2 & $6\times10^{3}$ & Bernal bilayer; Gated & \cite{Chen2026BLG} \\
\end{longtblr}
\endgroup

\clearpage

\renewcommand{\bibsection}{\section*{Supplemental References}}
\putbib[supplemental-references]
\end{bibunit}

\makeatletter
\auto@bib@empty
\makeatother
\end{document}